\documentclass{aastex}
\usepackage{emulateapj5}
\usepackage{latexsym}
\usepackage{amsmath}
\lefthead{NAKAMURA \& SHIGEYAMA}

\def\Msun{~M_{\odot} }
\def\Rsun{~R_{\odot} }
\def\cm3{{\rm ~cm}^{-3}}

\def\ltsima{$\; \buildrel < \over \sim\;$}
\def\ltsim{\lower.5ex\hbox{\ltsima}}
\def\gtsima{$\; \buildrel > \over\sim \;$}
\def\gtsim{\lower.5ex\hbox{\gtsima}}

\begin{document}
\title{Roles of Supernova Ejecta in Nucleosynthesis of Light Elements, Li, Be, and B} 

\author{Ko Nakamura  \& Toshikazu Shigeyama}

\affil{Research Center for the Early Universe, Graduate
School of Science, University of Tokyo, Bunkyo-ku, Tokyo 113-0033,
Japan}

\begin{abstract}
Explosions of type Ic supernovae (SNe Ic) are investigated using a relativistic hydrodynamic code to study roles of their outermost layers of the ejecta in light element nucleosynthesis through spallation reactions as a possible mechanism of the "primary" process. We have confirmed that the energy distribution of the outermost layers with a mass fraction of only $0.001$ \% follows the empirical formula proposed by previous work when the explosion is furious.  In such explosions, a significant fraction of the ejecta ( $>$0.1 \% in mass ) have the energy greater than the threshold energy for spallation reactions. On the other hand, it is found that the outermost layers of ejecta become more energetic than the empirical formula would predict when the explosion energy per unit ejecta mass is smaller than $\sim 1.3\times 10^{51}\mbox{ ergs/}\Msun$. As a consequence, it is necessary to numerically calculate explosions to estimate light element yields from SNe Ic. The usage of the empirical formula would overestimate the yields by a factor of $\gtsim 3$ for energetic explosions such as SN 1998bw and underestimate the yields by a similar factor for less energetic explosions like SN 1994I. The yields of light elements Li, Be, and B (LiBeB) from SNe Ic are estimated by solving the transfer equation of cosmic rays originated from ejecta of SNe Ic and compared with observations. The abundance ratios Be/O and B/O produced by each of our SNe Ic models are consistent with those of metal-poor stars. The total amounts of these elements estimated from observations indicate that energetic SNe Ic like SN 1998bw could be candidates for a production site of Be and B in the Galactic halo only when the fraction of this type out of all the SNe was a factor of $>100$ higher than the value estimated from current observational data.  This primary mechanism would predict that there are stars significantly deficient in light elements which were formed from the ISM not affected by SNe Ic. Since this has no support from current observations, other primary mechanisms such as the light element formation in superbubbles are needed for other types of SNe. The observed abundance pattern of all elements including heavy elements in metal-poor stars suggests that these two mechanisms should have supplied  similar amounts of Be and B. Our calculations show that SNe Ic can not produce an appreciable amount of Li.
\end{abstract}
\keywords{nucleosynthesis --- relativity --- shock waves --- cosmic rays
 --- supernovae: general}

\section{Introduction}
	The amounts of the light elements Li, Be, and B (LiBeB) at present are thought to be the sum of the products of the big bang nucleosynthesis and the subsequent cosmic-ray spallation reactions. The contribution from the big bang nucleosynthesis to $^7$Li is thought to be most pronounced among these three elements. Though some fractions of Li have been depleted inside some cool stars, its abundances on the surfaces of metal-poor stars are the only information that we can use to  identify the contribution from the big bang nucleosynthesis \citep[e.g.,][]{Ryan_99}. Thus it is important to know the contribution from the cosmic-ray spallation reactions to Li in the early stages of the Galaxy evolution during which metal-poor stars were formed. With this knowledge, we can constrain the cosmological parameters to synthesize light elements in the big bang nucleosynthesis \citep{Ryan_00}. 
	
	In addition, the investigation of the evolution of these elements in the early  stages of the Galaxy enables us to understand the origin of  cosmic-rays responsible for the light element nucleosynthesis. Recent observations of extremely metal-poor stars  apparently suggest that the primary (not secondary) spallation process dominated the light element nucleosynthesis in the early Galaxy \citep{Duncan_92}. In other words, cosmic-rays composed of C/O interacting with protons and/or He nuclei in the interstellar medium (ISM) have predominantly produced LiBeB.  A primary mechanism, in which LiBeB were produced in supernova ejecta-enriched superbubbles, was suggested to explain the Be evolution in the early Galaxy \citep{Higdon_98}. Since then, two-component models in which light elements are produced by standard Galactic cosmic-rays and metal-enriched particles in superbubbles have been investigated by several authors \citep{Ramaty_00, Fields_00, Suzuki_01}. These studies concluded that a primary mechanism is needed to explain the observed abundance trends of Be and B with O and Fe. 
Recently,  \citet{Suzuki_01} investigated the chemical evolution of LiBeB in the early epoch of the Galaxy using an inhomogeneous chemical evolution model developed by \citet{Tsujimoto_99} in which star formation is assumed to be induced by supernova (SN) explosions. Their model succeeded in reproducing not only the observed metallicity distribution of Galactic halo stars but also the observed abundance correlations of heavy elements. \citet{Suzuki_01} also considered two origins of Galactic cosmic-rays that synthesize light elements. One is from the ISM accelerated by SN shocks, and the other from SN ejecta accelerated by the SN shock. The authors claimed that $\sim$2 \% of Galactic cosmic-rays must be originated from the SN ejecta to reproduce the behavior of the abundance of Be at the metal-poor ends. However, their model suffers from shortage of energy supply from each SN to cosmic-rays. Moreover, the energy distributions of the cosmic-rays from these two origins might be different, though \citet{Suzuki_01} assumed they are the same.

	\citet{Fields_96} found from a numerical model of type Ic supernovae (SNe Ic) \citep{Nomoto_90} that the outermost C/O layers of  an SN could attain energies beyond the threshold value to produce LiBeB ($\sim$30 MeV per nucleon for O$+$H$\rightarrow$Be, which corresponds to the Lorentz factor of $\sim$1.03). Energetic explosions of massive stars with  stripped H-rich and He layers might be able to produce cosmic-rays enriched with C/O in the outermost layers of ejecta.  In the model of \citet{Nomoto_90}, the outermost layers of the ejecta do not become so relativistic. It may, however, be attributed to their coarse zoning in the outermost layers, which has to be as accurate as possible for this purpose.  Furthermore, their numerical calculations were performed with a hydrodynamic code that does not take into account relativistic effects. Thus it was impossible to accurately estimate the contribution from SNe Ic to the light element nucleosynthesis with their results.

The phenomena taking place when the supernova blast wave hits the surface of a relatively compact star were investigated in a more sophisticated fashion by \citet{Ensman_92} and \citet{Blinnikov_00} with their radiation-hydrodynamic codes. Their codes allow the radiation and the gas to go out of equilibrium. Both of their codes were non-relativistic except that they included light-travel-time corrections. They were concerned with the shock breakout of SN 1987A, because the most detailed observations immediately after the shock breakout were available for this SN. One of their main objectives was a detailed modelling of the UV burst immediately after the shock emergence. In addition to these numerical approaches, there have been semi-analytic approaches to the shock emergence in the plane-parallel medium in which the flow is assumed to be self-similar \citep{Gandel'man_56, Sakurai_60}. \citet{Kazhdan_92}  took into account the sphericity of the flow in a neighborhood of the surface. These semi-analytic approaches are also limited to the non-relativistic flow.

After the emergence of a very bright type Ic supernova, SN 1998bw, was found to be associated with a $\gamma$-ray burst GRB 980425 \citep{Galama_98} and \citet{Kulkarni_98} inferred from their radio observations that the radio shock front of SN 1998bw was moving at relativistic speeds with the Lorentz factor between 1.6 and 2, the relativistic motion from supernova explosions has been investigated. \citet{Matzner_99} estimated how much mass of relativistic ejecta could be obtained from the explosion of a massive star and derive an empirical formula giving the mass of relativistic ejecta from explosions of stars with simple polytropic density structures. Later, \citet{Tan_01} revised the empirical formula using their relativistic hydrodynamic code. Their formula would give $\sim 10^{-6}\,M_\odot$ of relativistic ejecta (Lorentz factor $>2$) from the explosion of a star with a mass of $10\,M_\odot$ and an energy of $10^{52}$ ergs.

Using the empirical formula of \citet{Matzner_99} for the energy distribution of SN ejecta,  \citet{Fields_02} concluded that SNe Ic, especially energetic events like SN 1998bw significantly contribute to light element nucleosynthesis through spallation reactions. The energy distribution of particles in SN ejecta used in \citet{Fields_02} follow a power law with a power index of $-4.6$. This power index is quite different from that of the energy distribution of the ISM accelerated by SN remnant shock \citep[e.g.][]{Meneguzzi_71}. Then \citet{Fields_02}  adopted a ``thick target''  approximation instead of solving the cosmic-ray transfer equation to derive the yields of LiBeB from the energy distribution of SN ejecta.

There still remain some problems to be addressed in their work;  the model for stellar structures \citet{Matzner_99} assumed is so simple that it involves suspicions of inaccurate estimates. In addition, they set the adiabatic index $\gamma$=constant over the whole stages of explosions, which may vary according to the physical conditions and affect the resultant energy distribution of the ejecta.

Our objective is to construct a realistic model for SN ejecta moving at relativistic speeds and to investigate their contribution to light element syntheses.
To improve the above situations, we calculate the energy distributions of ejecta as a result of SN explosions using a relativistic hydrodynamic (RHD) code with realistic numerical models for massive stars as the initial conditions. The atmospheres in radiative equilibrium are attached to the original models for massive stars (see references in Table \ref{tbl-model}). This is essential to investigate the energy distribution of ejecta at highest energies after the shock breakout. We also verify the validity of $\gamma$=constant by using more realistic equation of states that incorporates radiation and ideal gas in thermodynamic equilibrium. To investigate the change of the energy distribution of relativistic ejecta transferring in the ISM, we solve the transfer equation that takes into account the energy loss due to  ionization and spallation reactions. From these calculations, the amounts of synthesized LiBeB through primary spallation reactions are obtained.
	
	In the next two sections, we describe our RHD code in  Lagrangian coordinates (\S2) and initial conditions (\S3). Then, we show the results in \S4,  compare them with that of other authors, and discuss the differences. In \S5, we estimate the yields of LiBeB using our explosion models together with the leaky box model.
	
\section{Numerical code for relativistic hydrodynamics}
We have constructed a numerical code for RHD in Lagrangian coordinates based on the formalism presented in \citet{Marti_96}. Our RHD code well reproduces the exact solutions of test problems presented in \citet{Marti_96}, such as relativistic shock tubes.

\subsection{Equations}
In Eulerian coordinates, equations of RHD of a perfect fluid can be written in the following form:
\begin{equation}\label{mass}
 \frac{\partial D}{\partial t} + \nabla \cdot (D\mathbf{v}) = 0,
\end{equation}
\begin{equation}\label{momentum}
\frac{\partial \mathbf{S}}{\partial t} + \nabla (\mathbf{S} \cdot \mathbf{v} + p) =0,
\end{equation}
\begin{equation}\label{energy}
\frac{\partial \tau}{\partial t} + \nabla \cdot (\mathbf{S}-D\mathbf{v}),
\end{equation}
where $t$, $D$, $\mathbf{S}$, $\tau$ denote the time,  the rest-mass density, the momentum density, and the energy density in a fixed frame where the fluid moves at speed $\mathbf{v}$, respectively. These variables are related to quantities in the local rest frame of the fluid through
\begin{equation}
D = \rho W
\end{equation}
\begin{equation}
\mathbf{S} = \rho h W^2 \mathbf{v}
\end{equation}
\begin{equation}
\tau = \rho h W^2 -p -D,
\end{equation}
where $\rho$, $p$, $W$ and $h$ denote the proper rest-mass density, the pressure, the fluid Lorentz factor, and the specific enthalpy, respectively. The specific enthalpy is given by
\begin{equation}
h = 1 + \varepsilon + \frac{p}{\rho},
\end{equation}
where $\varepsilon$ is the specific internal energy. 

	We can describe these equations with a Lagrangian coordinate $m$. Since we are concerned with spherically symmetric explosions, we will rewrite them in the 1-dimensional spherical coordinate system. \begin{equation}\label{massl}
\frac{dV}{dt} - 4\pi \frac{\partial(r^2 v)}{\partial m} = 0,
\end{equation}
\begin{equation}\label{momentuml}
\frac{ds}{dt} + 4\pi r^2 \frac{\partial p}{\partial m} +\frac{G M_r}{r^2}= 0,
\end{equation}
\begin{equation}\label{energyl}
\frac{dQ}{dt} + 4\pi \frac{\partial (r^2 p v)}{\partial m} = 0,
\end{equation}
where $v$ denotes the radial velocity, $G$ the gravitational constant, $M_r$ the mass included within the radius $r$, $V=1/D$, $s=V\mathbf{S}\cdot\mathbf{r}/r$ ($\mathbf{r}$: the radial vector) and
\begin{equation}
Q=\tau V - \frac{G M_r}{r}.
\end{equation}
 The Lagrangian coordinate $m$ is related to $r$ through
\begin{equation}
\partial m = \frac{4\pi r^2}{V} \partial r.
\end{equation}
The gravity is included in the weak limit where no general relativistic effect is prominent. The pressure is described as the sum of the radiation pressure and the gas pressure:
\begin{equation}\label{eqn-eos}
p = \frac{a}{3} T^4 + \frac{k \rho T}{\mu m_p},
\end{equation}
where $a$ denotes the radiation constant, $T$ the temperature, $k$ the Boltzmann constant, $\mu$ the mean molecular weight, and $m_p$ the mass of proton. The specific internal energy $\varepsilon$ is given by
\begin{equation}
\varepsilon = \frac{a T^4}{\rho} + \frac{3kT}{2 \mu m_p}.
\end{equation}
Here we will introduce two kinds of adiabatic indices, $\gamma_1$ and $\gamma_2$ defined by
\begin{equation}
\gamma_1 = \left( \frac{d \ln p}{d \ln \rho} \right)_S,
\end{equation}
\begin{equation}
\gamma_2 = \frac{p}{\rho \varepsilon} + 1.
\end{equation}

The difference scheme of equations (\ref{massl})--(\ref{energyl}) can be written in the following form;
\begin{equation}\label{deff2}
\begin{split}
		\begin{pmatrix}    V^{n+1}_i \\ s^{n+1}_i \\ Q^{n+1}_i    \end{pmatrix}
		 =
		\begin{pmatrix}    V^{n}_i \\ s^{n}_i \\ Q^{n}_i 		  \end{pmatrix}
		&+\frac{4\pi\varDelta t}{\varDelta m}
		\begin{pmatrix}    -(r^2 v)_{i-1/2} + (r^2 v)_{i+1/2} \\
					(r^2p)_{i-1/2} - (r^2p)_{i+1/2} \\
					(r^2 p v)_{i-1/2} - (r^2 p v)_{i+1/2}
		\end{pmatrix}  \\
		&+ \varDelta t
		\begin{pmatrix}    0 \\ - \frac{G M_r}{r^2_i} \\ 0 		  \end{pmatrix}.
\end{split}
\end{equation}
We use the Godunov method to numerically solve equations (\ref{deff2}) distinguishing these two adiabatic indices in our Riemann solver. 

The number of zones are listed in Table \ref{tbl-model}.  To obtain accurate energy distributions of SN ejecta above the threshold energies for cosmic-ray spallation reactions, our models have zones in the outermost layers with masses well below $10^{-9}\ \Msun$.

\subsection{Boundary Conditions}

	When we calculate the explosion of a star, we need to set the boundary conditions at the outer edge of the star as follows;
\begin{equation}
	\begin{matrix} p_{imax+1/2} = 0, \\ v_{imax+1/2} = v_{imax}, \end{matrix}
\end{equation}
where $imax$ is the zone number corresponding to the outermost layer. 

At the center, which corresponds to the zone interface $i=1/2$, we consider an imaginary zone $i=0$, where the physical quantities are given by
\begin{equation}
	\begin{matrix} p_0 = p_1,\,  \rho_0 = \rho_1 \\ v_0 = -v_1,\,r_0=-r_1, \end{matrix}
\end{equation} 
to keep the symmetry with respect to the center.
\section{Initial conditions}

	We consider four stellar models immediately before the core collapse as the initial conditions. These stars are  originated from $12 - \sim 40\Msun$ main-sequence stars. Three of them are thought to have undergone  intense stellar winds and lost their H-rich and He envelopes. As a result, the stellar surfaces mainly consist of carbon and oxygen at explosion.
	These three models have corresponding real type Ic supernovae as shown in Table \ref{tbl-model}. These stars have become fairly compact with radii less than the solar radius. This compactness results in higher pressures at the shock breakout compared with explosion of  a star with an extended envelope if the explosion energies per unit ejecta mass are the same. The other star is thought to be the progenitor of SN 1987A, one of the best studied supernovae. We have calculated the explosion for this supernova and compare the result with the previous work to check our numerical code.

\begin{table*}[htb]
\begin{center}
\caption{Parameters of stellar models.}
\begin{tabular}{crrrrccll}
\hline
$M_{ms}$\tablenotemark{a} & $M_*$\tablenotemark{b}   & $R_*$\tablenotemark{c} 
& $M_{\rm ej}$ & $M_{\rm O}$\tablenotemark{d}  &
$E_{\rm ex}$ & Number of zones & SN\tablenotemark{e} &  Reference\\ 
$(\Msun)$ & $(\Msun)$ & $(\Rsun)$  & $(\Msun)$  & $(\Msun)$   & 
$(\times 10^{51}$ergs) &  &  &  \\ \hline

-40       & 15  & 0.29 & 13 & 10 & 30 & 389 & 1998bw & \cite{Nakamura_01}\\
20-25 & 4.6 & 0.32 & 2.9 & 2.1 &  4   & 398 & 2002ap  & \cite{Mazzali_02}\\
12-15 & 2.4 & 0.24 & 0.99 & 0.66 &1   & 294 & 1994I      & \cite{Iwamoto_94}\\ \hline
\tablenotetext{a}{The stellar mass on the main sequence.}
\tablenotetext{b}{The stellar mass at core collapse}
\tablenotetext{c}{The stellar radius at core collapse}
\tablenotetext{d}{The mass of oxygen in the ejecta}
\tablenotetext{e}{Corresponding real supernova }
\end{tabular}
\end{center}
\label{tbl-model}
\end{table*}

	To initiate explosions, we first replace the central Fe core with the point mass located at the center,  release the energy in the innermost several zones  in the form of thermal energy (or pressure), and trace the evolution of physical quantities. The calculations are stopped when the ejecta keep expanding homologously. The parameters of each model are tabulated in Table \ref{tbl-model}.
	
\section{The energy distribution of supernova ejecta}
\subsection{SN 1987A}
To check our numerical code, we have performed a calculation for a type II SN 1987A and compare the resultant velocity distribution as a function of enclosed mass  with that calculated by \citet{Shigeyama_90} with a non-relativistic hydrodynamic code. Results of these two calculations are found to be in  agreement  to a fairly good accuracy ($\ltsim$0.2 \% on average).

\subsection{SN 1998bw}
The explosion energy has been found to be considerably higher than 10$^{51}$ ergs from modelling the observed light curve and spectra of SN 1998bw. Asphericity has been suggested from the observed feature of emission lines of O and Fe \citep{Maeda_02}. The fact that spherical models cannot explain the light curve both in the early and late phases also suggests an aspherical explosion : The spherical explosion with an energy of $\sim5\times 10^{52}$ ergs can explain the early phases but the light curve was predicted to decline too fast to be compatible with the observations, while that with a lower energy $\sim 7\times 10^{51}$ ergs cannot evolve as fast as observed in the early phase but can explain the observed decline of the late light curve \citep{Nakamura_01}.

\begin{figure*}[ht]
\begin{center}
\plotone{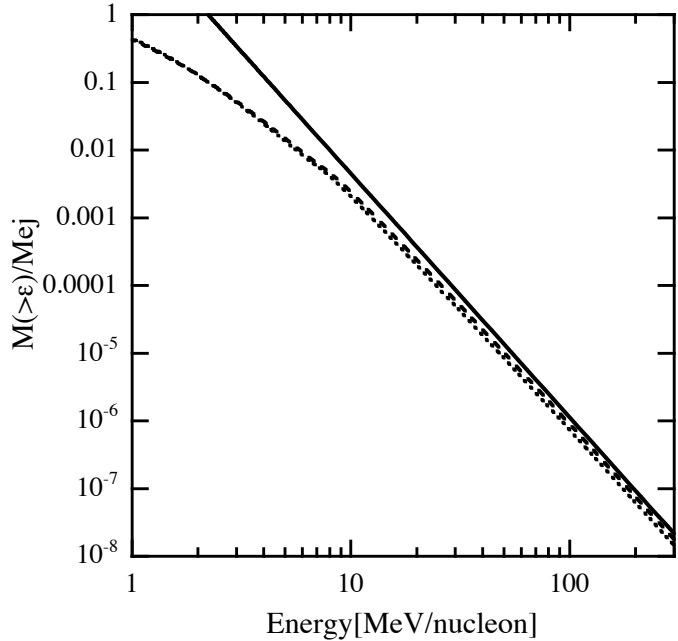}
	\caption{The integrated energy distribution $M(>\epsilon)$ of  ejecta for SN 1998bw model with $E_{\rm ex}=3\times 10^{52}$ ergs. The solid line represents the fitting formula of \citet{Fields_02}. The results of our numerical calculation with  the constant adiabatic index  ($\gamma_1=\gamma_2=4/3$) is shown by the dotted line. The dashed line shows the result with  the variable adiabatic index.}
	\label{fig-fieldsa}
\end{center}
\end{figure*}

Here we have calculated an explosion of a $\sim15\,\Msun$ C/O star with an explosion energy $E_{\rm ex}$ of $3\times 10^{52}$ ergs for this energetic supernova (Table \ref{tbl-model}). The mass of the ejecta $M_{\rm ej}$ becomes 13$\Msun$ with the rest of $\sim 2\,\Msun$ collapsed to become a compact object at the center. The resultant energy distributions of the ejecta are plotted in Figure \ref{fig-fieldsa} with  that predicted from the empirical formula proposed by \citet{Fields_02}:
\begin{equation}\label{eqn-fields}
 \begin{split}\frac{M(>\epsilon)}{M_{\rm ej}}= &
2.2\times 10^{-4}\left(\frac{E_{\rm ex}}{10^{51}\,\mbox{ergs}}\right)^{3.6}\left(\frac{M_{\rm ej}}{1\, \Msun}\right)^{-3.6} \\ &\times\left(\frac{\epsilon}{10\,\mbox{MeV}}\right)^{-3.6}, \end{split}
\end{equation}
where $M(>\epsilon)$ denotes the mass of ejecta that have particle energy per nucleon greater than $\epsilon$. 
At the high energy tail, the energy distribution from our calculation with realistic equation of states (the dashed line in Fig. \ref{fig-fieldsa}) show the same dependence on energies as the above empirical formula but with somewhat ($\sim 13$ \%) smaller amounts of matter. The adiabatic indices ($\gamma_{1}$ and $\gamma_{2}$) fixed to 4/3 also agree with the empirical formula (the dotted line in Fig. \ref{fig-fieldsa}). Thus this agreement is ascribed to the features of this particular model that the density distribution of the star is well approximated by a simple formula presented in \citet{Matzner_99} and that the adiabatic indices in the shocked ejecta are close to 4/3.
 At around the threshold energies for spallation reactions ($\sim 5$--$\sim 30$ MeV/nucleon), the energy distribution from our calculation significantly deviates from equation (\ref{eqn-fields}) (see Fig. \ref{fig-fieldsa}). Therefore, it is doubtful to use equation (\ref{eqn-fields}) to estimate yields of light elements synthesized by the spallation reactions. The explosion of this supernova is so furious that the assumption of $M(>\epsilon)/M_{\rm ej}<<1$ already breaks down in this energy region. In summary, equation (\ref{eqn-fields}) always overestimates the mass with given $\epsilon$. In particular, it overestimates the mass with $\epsilon$ greater than the threshold energies by a factor of $\gtsim 3$.

\subsection{Other type Ic supernovae}
\begin{figure*}[ht]
\plotone{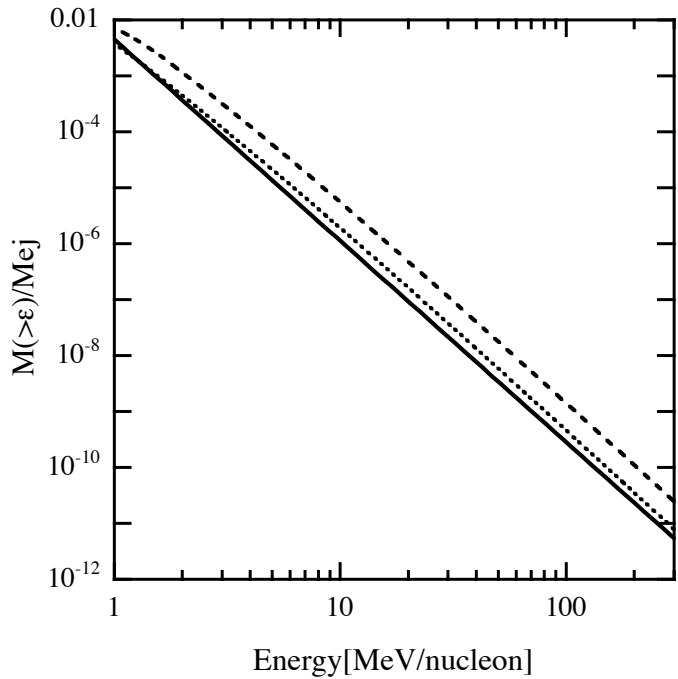}
	\caption{The same as Fig.  \ref{fig-fieldsa} but for $E_{\rm ex}=3 \times 10^{51}$ ergs. Our result with the variable adiabatic index indicates more efficient accelerations compared with the fitting formula of \citet{Fields_02}}.
	\label{fig-fieldsb}
\end{figure*}

\begin{figure*}[ht]
\plotone{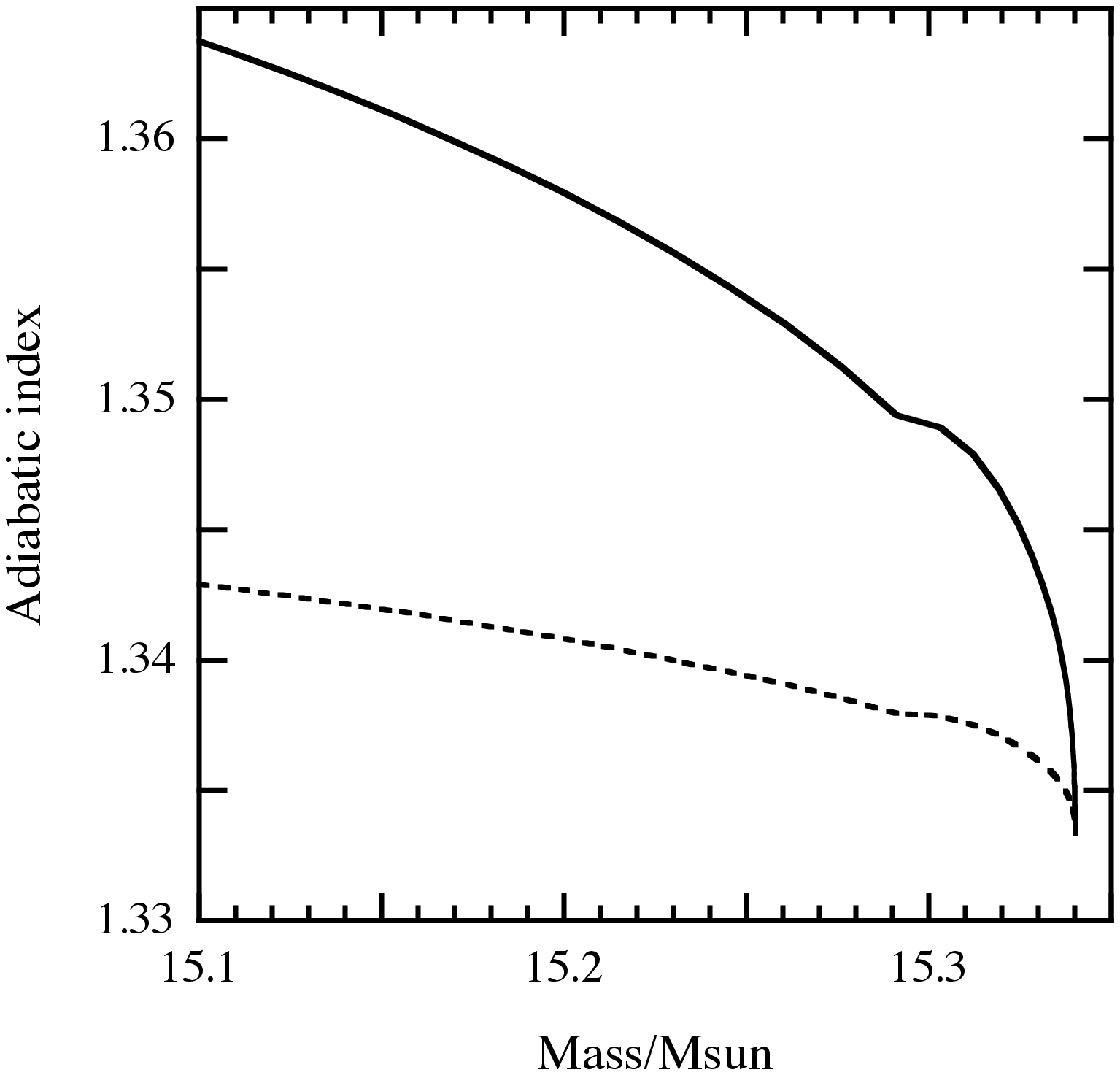}
	\caption{The distributions of adiabatic index $\gamma_1$ as a function of the enclosed mass (including the central remnant mass) immediately after the shock breakout for the SN 1998bw models. The solid line shows the model with $E_{\rm ex} = 3 \times 10^{51}$ ergs. The dashed line with $E_{\rm ex} = 3 \times 10^{52}$ ergs.}
	\label{fig-ai}
\end{figure*}

The other type Ic supernovae listed in Table \ref{tbl-model} are explosions with explosion energies per unit mass smaller than SN 1998bw model. The energy distribution of the ejecta of these supernovae follow a power law with the same index in equation (\ref{eqn-fields}) but with a significantly greater amount of matter accelerated to given energies per nucleon. A low $E_{\rm ex}/M_{\rm ej}$ leads to adiabatic indices in the shocked matter greater than 4/3. The stiffer equation of states leads to a more efficient acceleration. The same is true for a model of SN 1998bw with a less explosion energy (e.g. $3\times 10^{51}$ ergs) as shown in Figures \ref{fig-fieldsb} and \ref{fig-ai}. As a result, equation  (\ref{eqn-fields}) underestimates the mass of ejecta with given $\epsilon$ for these supernovae.
	
\subsection{Energy distribution with a variable adiabatic index}
The energy distributions ($-dM(>\epsilon)/d\epsilon$) for all the models have smaller power-law indices ($\sim -$4.6) than that of the cosmic-rays accelerated through the shock-fronts of SN remnants $\sim -$2.6 \citep[e.g.][]{Meneguzzi_71}. This will affect the ratios of light elements synthesized from the cosmic-ray spallation reactions (see \S\ref{sect-yields}). 

From the discussion in the preceding sections, it follows that equation (\ref{eqn-fields}) holds when the adiabatic index is very close to 4/3 and $M(>\epsilon)/M_{\rm ej}\ltsim 10^{-5}$. In particular, our calculations show that the energy distribution does not become a simple power-law for  $M(>\epsilon)/M_{\rm ej}\gtsim 10^{-5}$. This must be due to the fact that the simple stellar density distribution \citet{Matzner_99} assumed can reproduce the density distributions in realistic stellar model only for the outer 0.001 \% layers in mass. When the explosion energy is furious, the adiabatic index is close to 4/3 but the mass with $\epsilon$ greater than the threshold energies for spallation reactions can become significantly greater than $10^{-5}M_{\rm ej}$ as in SN 1998bw model. Thus equation (\ref{eqn-fields}) does not give a good estimate for the mass of ejecta in the region between $\epsilon$ and $\epsilon+d\epsilon$ around the threshold energies ($-dM(>\epsilon)/d\epsilon$) as well as $M(>\epsilon)$.

The smaller the explosion energy per unit mass $E_{\rm ex}/M_{\rm ej}$ becomes, the more significant our results deviate from equation (\ref{eqn-fields}) even in the high energy tails (Fig. \ref{fig-fieldsb}) .
Our calculations for all the models listed in Table \ref{tbl-model} with a constant adiabatic index equal to 4/3 always result in the energy distributions of ejecta in good agreement with their fitting formula at high energy tails irrespective of values of  $E_{\rm ex}/M_{\rm ej}$. Therefore this deviation should be caused by variable adiabatic indices. At the same time, the agreement of these results with equation (\ref{eqn-fields}) suggests that the simple density distributions used in \citet{Matzner_99} are good approximations of realistic stars in the outermost layers (0.001 \% of the ejecta mass).

\begin{figure*}[hb]
\plotone{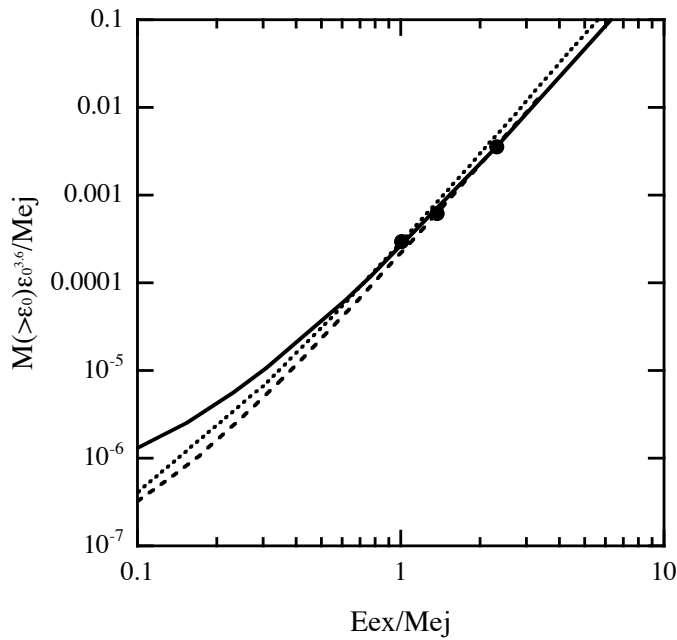}
	\caption{$M(>\epsilon_0) \times \epsilon_0^{3.6} / M_{\rm ej}$ as a function of $E_{\rm ex}/M_{\rm ej}$ where $\epsilon_0$ = 100 MeV/nucleon. $M(>\epsilon_0)$ and $M_{\rm ej}$ are in units of $\Msun$, and $E_{ex}$ in $10^{51}$ergs. The solid, dashed, and dotted lines show models for SN 1998bw,  SN 2002ap, and SN 1994I, respectively. The values of $M(>\epsilon) \times \epsilon^{3.6}$ for  $\epsilon \gtrsim $ 10 MeV/nucleon are almost constant for different models with the same value of $E_{ex}/M_{ej}$ as long as $E_{ex}/M_{ej}>6\times 10^{50}\,\mbox{ergs}/\Msun$.  The filled circles correspond to $E_{\rm ex}$ shown in Table \ref{tbl-model}.}
	\label{fig-fit}
\end{figure*}
From our calculations with a realistic equation of state (Eq. (\ref{eqn-eos})) varying explosion energies for three SN models in Table \ref{tbl-model}, we have plotted $M(>\epsilon)\times\epsilon^{3.6} / M_{\rm ej}$ at $\epsilon=100$ Mev/nucleon as a function of $E_{\rm ex}/M_{\rm ej}$ in Figure \ref{fig-fit}. For less energetic explosions $E_{\rm ex}/M_{\rm ej}\ltsim 6\times10^{50}$ ergs$/\Msun$, the lines of the three models deviate from each other. Thus a single fitting formula such as equation (\ref{eqn-fields}) cannot account for all the realistic explosions. We need three different parameters $A$ for these three SN models to express the energy distribution of ejecta as
\begin{equation}\label{eqn-fit}
 	\begin{split}
\frac{M(>\epsilon)}{M_{\rm ej}}= & A \left(\frac{E_{\rm ex}}{10^{51}\,\mbox{ergs}}\right)^{3.4}\left(\frac{M_{\rm ej}}{1\, \Msun}\right)^{-3.4} \\ &\times\left(\frac{\epsilon}{10\,\mbox{MeV}}\right)^{-3.6}, \end{split}
\end{equation}
that takes into account variable adiabatic indices originated from the equation of states  (Eq. (\ref{eqn-eos})). The constant $A$ is equal to $1.9 \times 10^{-4}$ for SN 1998bw model, $2.1 \times 10^{-4}$ for SN 2002ap, and $2.8 \times 10^{-4}$ for SN 1994I.

\section{Light element nucleosynthesis}
To investigate the role of supernova ejecta in light element nucleosynthesis, the modification of the energy distribution of ejecta when they transfer in the ISM need to be considered because the cross sections of spallation reactions are sensitive to energies of particles.

\subsection{Transfer equation}
Energetic ejecta accelerated by a supernova explosion lose energy when they collide with neutral atoms in the ISM and ionize them. The ionization energy loss rate of element $i$  in the ejecta through H gas, $\omega_i$ (MeV/s), is given  by the formula \citep{Schlickeiser_02}:
\begin{equation}
	\begin{split}
		\omega_i = & Z_{\mathrm{eff},i}^2 \times 1.82 \times 10^{-13} n_{\rm H\,I}/A_i         \\[-3pt]
				  & \times \bigl\{ 1 + 0.0185 \ln \beta H(\beta - 0.01) \bigr\} 
		  		       \frac{2 \beta^2}{10^{-6} + 2 \beta^3},
	\end{split}
\end{equation} 
where $n_{\rm H\,I}$ is the number density of neutral H in the ISM, $\beta = v/c$ and $H(x)$ denotes the Heaviside step function. The effective charge $Z_{\mathrm{eff},i}$ is expressed as
\begin{equation}
	Z_{\mathrm{eff},i} = Z_i \bigl\{ 1-1.034 \exp\left(-137\beta Z_i^{-0.688}\right) \bigr\},
\end{equation}
where $Z_i$ is the atomic number of the element $i$ in the ejecta.

	We use the leaky-box model \citep{Meneguzzi_71} and the transfer equation for the mass of the  element $i$ with an energy per nucleon $\epsilon$ at time $t$, $F_i (\epsilon,t)$, is expressed as
\begin{equation}\label{eqn-transfer}
	\begin{split}
		\frac{\partial F_i(\epsilon,t)}{\partial t} = & \frac{\partial [\omega_i (\epsilon) F_i(\epsilon,t)]}{\partial \epsilon} \\
		  &- \left(\frac{F_i(\epsilon,t)}{\Lambda_{\rm esc}} + \frac{F_i(\epsilon,t)}{\Lambda_{n,i}} \right) \rho v_i(\epsilon),
	\end{split}
\end{equation}
where $\Lambda$'s are the loss lengths in g $\mathrm{cm}^{-2}$, $\rho$ denotes the mass density of the ISM, $v_i(\epsilon)$ the velocity of the element $i$ with an energy per nucleon of $\epsilon$. $\Lambda_{\rm esc}$ denotes the range before escaping from a given system (we assume $\Lambda_{\rm esc}=100$ g cm$^{-2}$ following \citet{Suzuki_01}), and $\Lambda_{n,i}$ due to spallation reactions. The latter is  expressed as  
\begin{equation}
	\Lambda_{n,i}(\epsilon)=\frac{n_p m_p + n_\mathrm{He} m_\mathrm{He}}{n_p \sigma_{p,i}(\epsilon) + n_\mathrm{He} \sigma_{\mathrm{He},i}(\epsilon)}.
\end{equation}
Here the total cross sections of spallation reactions between  particle $i$ and proton or He are denoted by $\sigma_{p,i}$ or $\sigma_{\mathrm{He},i}$, respectively. The mass of He is denoted by $m_{\rm He}$. The number densities of proton and He in the ISM have been introduced as $n_p$ and $n_{\rm He}$ and we assume that the ISM is uniform and neutral: $n_p = n_{\rm H\,I}=1\,\mathrm{cm}^{-3}$ and $n_\mathrm{He} = 0.1\,\mathrm{cm}^{-3}$. The ionization energy loss rate $\omega_i$ is proportional to $n_{\rm H\,I}$. The energy loss rate is, however, not so sensitive to the ionization state of H,  since the energy loss rate through Coulomb collisions with free electrons in the ionized medium is comparable to that in a neutral medium and have a similar energy dependence.

 Equations (\ref{eqn-transfer})  for C and O are numerically solved with initial conditions given by the energy distribution presented in the previous section:
\begin{equation}
F_i(\epsilon,0) = -X_i(\epsilon)\frac{dM(>\epsilon)}{d\epsilon}.
\end{equation}
Here $X_i(\epsilon)$ denotes the mass fraction of the element $i$ with an energy per nucleon of $\epsilon$. Equations (\ref{eqn-transfer}) are implicitly integrated with respect to time. 

\begin{figure*}[ht]
\plotone{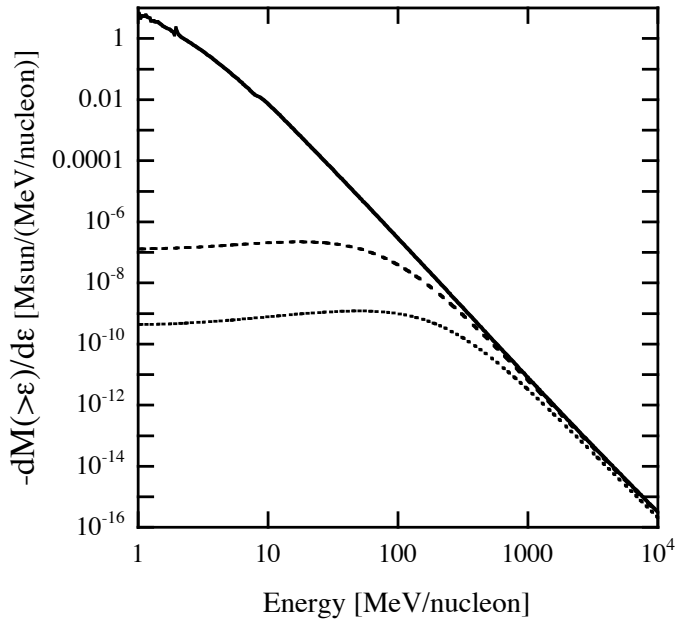}
	\caption{The attenuation of the energy distribution of O for SN 1998bw model with $E_{\rm ex}=3 \times 10^{52}$ ergs is shown. The solid line is the initial energy distribution  of the ejecta in the homologous expansion phase.  The dashed line shows the energy distribution attenuated by ionizations at $t=1$ Myr and the dotted line at $t=5$ Myr.}
	\label{fig-trans}
\end{figure*}
	Through the ionization energy loss, the energy distribution of ejecta becomes harder as time passes as shown in Figure \ref{fig-trans}. The effect of $\Lambda_{\rm esc}$ is negligible in this calculation. Ionization reduce the energies of most particles below the threshold energies for the spallation reactions before the particles escape from the system.

\begin{figure*}[hb]
\plotone{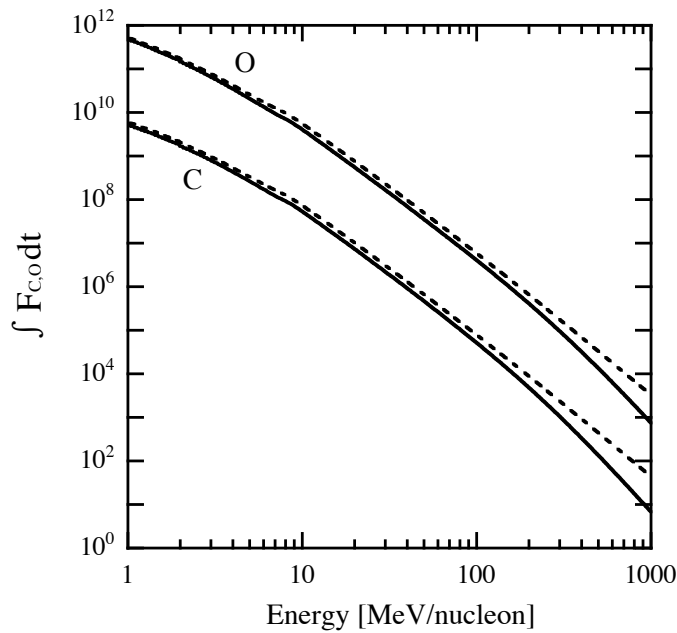}
	\caption{The solid lines show the time integrated energy distributions of C and O in the ejecta derived from direct integration of the transfer equation (\ref{eqn-transfer}). The dashed lines show the same quantities obtained with the ``thick target" approximation (Eq. (\ref{eqn-thicktarget})).}
	\label{fig-spectra}
\end{figure*}

Since \citet{Fields_02} used the ``thick target" approximation to calculate yields of light elements synthesized by cosmic-ray spallation reactions, we compare the energy distribution calculated from equation (\ref{eqn-transfer}) with that obtained from the ``thick target" approximation \citep{Ramaty_75}. This approximation assumes the time-integrated energy distribution of cosmic-rays as 
\begin{equation}\label{eqn-thicktarget}
\int_0^\infty F_i(\epsilon,t) dt  = \frac{R_i(\epsilon)X_i(\epsilon)}{\rho v_{i}(\epsilon)}\frac{M(>\epsilon)}{\epsilon},
\end{equation}
where the range of element $i$, $R_i(\epsilon)$, is defined as $R_i(\epsilon)=\rho v_i(\epsilon)\epsilon/\omega_i$. This relation can be derived from the integration of equation (\ref{eqn-transfer}) with respect to time $t$ and energy $\epsilon$ omitting the terms including $\Lambda$'s. The right hand side of this equation is plotted in Figure \ref{fig-spectra} with the dashed lines for C and O together with the time-integrated energy distribution obtained by solving equation (\ref{eqn-transfer}) (the solid lines). It is clear that this approximation gives a fairly good estimate for  the time-integrated energy distribution at around threshold energies (several to 30 MeV per nucleon). However, \citet{Fields_02} adopted a somewhat different approximation in which 
\begin{equation}
\int_0^\infty F_i(\epsilon,t) dt  = -\frac{R_i(\epsilon)X_i(\epsilon)}{\rho v_{i}(\epsilon)}\frac{dM(>\epsilon)}{d\epsilon}.
\end{equation}
This will lead to a factor of $\sim 3.6$ greater amount of yields of light elements. 

\subsection{Yields of light elements}\label{sect-yields}
	Light elements (LiBeB) produced through the spallation reactions; C, O $+$ H, He $\rightarrow ^{6,7}$Li, $^9$Be, $^{10,11}$B are considered here. The yield of a light element $l$ via the $i+j \rightarrow l+\cdots$ reaction is estimated by the following formula;
\begin{equation}
	\frac{dN_l}{dt} = n_j \int_{\epsilon_{\rm min}}^{\epsilon_{\rm max}} \sigma_{i,j}^l(\epsilon) \frac{F_i(\epsilon,t)}{A_i m_p} v_i(\epsilon) d\epsilon,
\end{equation}
where $N_l$ is the number of the produced light element $l$, $n_j$ is the number density of interstellar element $j$, $\sigma_{i,j}^l$ the cross section of $i+j \rightarrow l+\cdots$ reaction, $A_i$ the mass number of the element $i$.  The integration with respect to energy is carried out between the minimum and maximum energies per nucleon in the ejecta obtained from our numerical calculations, which are denoted as $\epsilon_{\rm min}$ and $\epsilon_{\rm max}$, respectively. We use the empirical formula given by \citet{Read_84} for the cross sections.

\section{Results and Comparison with observations}
\subsection{Abundance ratios}
Table \ref{tbl-yields} summarizes the results. Since the ratios of elements are determined solely by the energy distribution of the injected materials and the energy distributions of all the calculated models have a similar shape, the models give similar isotopic ratios: $^7$Li/$^6$Li$\sim$1.2 and $^{11}$B$/^{10}$B$\sim 2.8$. The B isotopic ratio observed in meteorites is 4.05$\pm0.05$ \citep{Shima_62, Chaussidon_95}. Thus neutrino spallation processes in core-collapse SNe may contribute to the enhancement of $^{11}$B . On the other hand, this ratio observed in the local ISM is around $3.4\pm 0.7$ \citep{Lambert_98}, which can be marginally reproduced by the mechanism discussed here. 

The element ratio B/Be$\sim 15$ from our models can explain this ratio on the surfaces of some halo stars because the average ratio of halo stars is thought to be 17$-$21 and some halo stars like HD94028 show fairly low ratio as $\sim 12$. The solar value 29.6 suggests that at least B in the solar system must have been synthesized in different sites in addition to neutrino processes.

\subsection{The amounts of Be and B in the Galactic halo}
Since the big bang nucleosynthesis is not expected to produce observable amounts of B or Be, it will be easier to identify which primary process is responsible for the enrichment of these elements as compared with Li.
The element abundances of Be are determined for more stars than the other light elements Li and B. The observed mass ratios of $X_{\rm Be}/X_{\rm O}$ on the surfaces of metal-poor stars range from $\sim10^{-9}$ to $\sim2\times10^{-8}$ \citep{Boesgaard_99}. The result presented in Table \ref{tbl-yields} is in good agreement with this observation. This is in contrast with the result of \citet{Fields_02}. They obtained $X_{\rm Be}/X_{\rm O}\sim10^{-6}$ for a furious SN Ic like SN 1998bw. This large amount of Be must be diluted by a large number of other SNe that do not supply such a large amount of Be to be reconciled with the observed $X_{\rm Be}/X_{\rm O}$ ratios. They found that the fraction of furious SNe Ic needs to be  less than $8\times 10^{-4}$ out of all the SNe. \citet{Ramaty_00} had proposed another primary process in which  SNe in superbubbles yield  $X_{\rm Be}/X_{\rm O}\sim 3\times 10^{-8}$ on average. As a consequence, the overproduction of Be from furious SNe ironically made \citet{Fields_02} to suggest that the supply of Be from furious SNe needs to be a tiny fraction of that from SNe in superbubbles. 

\begin{table*}[ht]
\begin{center}
\caption{Yields of light elements.}
\begin{tabular}{lrrrrrc} \hline
 &$^6$ Li   & $^7$ Li  & $^9$ Be &
$^{10}$ B  & $^{11}$ B & $\log M_{\rm Be}/M_{\rm O}$\\
Model & $(10^{-7}\,\Msun)$   & $(10^{-7}\,\Msun)$ & $(10^{-7}\,\Msun)$ &
$(10^{-7}\,\Msun)$  & $(10^{-7}\,\Msun)$ &    \\ \hline

SN 1998bw & 2.38 & 3.31 & 0.999 & 4.38 & 13.4 & -8.0\\
SN 2002ap & 0.0841 & 0.114 & 0.0348 & 0.152 & 0.464 & -8.8\\
SN 1994I    & 0.0140 & 0.0190 & 0.00578 & 0.0253 & 0.0776 & -9.1\\ 
\hline
\end{tabular}
\end{center}
\label{tbl-yields}
\end{table*}

On the other hand, our result does not need to dilute Be with other SNe. Instead, the Be/O ratio on the surface of a metal-poor star need to be determined by a single SN as was suggested for heavy elements \citep{Shigeyama_98}. This might be possible if the tangled magnetic fields in the ISM trap accelerated SN ejecta inside the region where the SNR shock eventually swept up. This mechanism predicts that there must be a considerable number of metal-poor stars with essentially no light elements. These are the descendants of SNe other than SNe Ic. It is, however, likely that this conjecture has been ruled out by current observations. No star without light elements has been reported. There is also a defect in the argument of \citet{Fields_02}.  The abundance pattern of heavy elements determined by multiple SNe in superbubbles must be at odds with the observations that show a large scatter in the stellar abundance ratios especially for Fe-group elements and $r$-process elements \citep[e.g.,][]{McWilliam_95, Cayrel_04}. To reproduce the trends of all the elements observed for metal-poor stars, both of these two sites should have supplied comparable amounts of  Be.
The quantitative discussion needs a detailed chemical evolution model not only for light elements but for heavy elements, which is beyond the scope of this paper.

The observed mass ratio $X_{\rm Be}/X_{\rm H}$ has a value $\sim 10^{-11}$ at [Fe/H]$\sim-1.5$ \citep{Boesgaard_99} which is the maximum value of [Fe/H] attained in the Galactic halo. The Fe yield from core-collapse SNe of $2.1\times X_{Fe\odot}$ per unit SN mass \citep{Tsujimoto_97}  indicates that the yield of Be per unit supernova mass is 
\begin{equation}
\frac{M_{\rm Be}}{M_{\rm SN}}\sim5\times10^{-10},
\end{equation}
 in the halo. The corresponding values for the three models in Table \ref{tbl-yields} are $\sim 2.5\times10^{-9}$ (SN 1998bw), $\sim 1.7\times10^{-10}$ (SN 2002ap), and $\sim 4.8\times10^{-11}$ (SN 1994I).  Observations suggest that the fraction of furious SNe Ic is $\sim$1/700 of core-collapse SNe \citep{Podsiadlowski_04}. This fraction should have been a factor of $>100$ higher in the early stage of the Galaxy for furious SNe Ic to have been a major production site for Be. The element abundance ratios B/Be from the present models consistent with those for Galactic halo stars indicate that the same is true for B in the halo.

\subsection{Li in the Galactic halo}
The observed number ratios Li/H on the surfaces of metal-poor stars suggest that the big bang nucleosynthesis leads to Li/H$\sim 1.23^{+0.68}_{-0.32}\times 10^{-10}$ and have a significant correlation with the metallicity \citep{Ryan_00}. If this correlation is extrapolated to [Fe/H]$=-1.5$, then the ratio Li/H becomes $\sim 4.6\times 10^{-10}$. Therefore the amount of  Li produced in the Galactic halo was  $\sim 2.5 \times 10^{-9}M_{\rm H}$, where $M_{\rm H}$ denotes the mass of H. Thus an argument similar to the preceding section results in the yield of Li per unit supernova mass of
\begin{equation}\label{eqn-li}
\frac{M_{\rm Li}}{M_{\rm SN}}\sim1.2\times10^{-7}.
\end{equation}
The corresponding value of SN 1998bw model is $1.4\times 10^{-8}$ and the other models give less values. If this value in equation (\ref{eqn-li}) is taken at face value, SNe Ic cannot produce this amount of Li. However, there seem to be few stars with the metallicity range of $-2\ltsim$[Fe/H]$\ltsim -1$ for which the Li abundance is measured \citep{Ryan_00}. We need to wait for abundance information in this metallicity range to deduce a firm conclusion.

 At the low metallicity end ([Fe/H] $\sim -4$), the observed Li/O ratios $\sim 10^{-4}$ \citep{Ryan_00} are more than two orders of magnitude greater than that in our SN 1998bw model (see Table \ref{tbl-yields}). The Li produced from this primary mechanism is likely to be negligible compared to the primordial Li.

\section{Conclusions and discussion}
We have performed numerical calculations for SNe Ic explosions using  a relativistic hydrodynamic code to investigate how much mass of ejecta is accelerated beyond the threshold energy for spallation reactions to synthesize light elements. In our calculations, realistic massive star models are used as the initial conditions of SNe and the EOS takes into account the thermal radiation and ideal gas. We have compared the resultant energy distributions of ejecta with the empirical formula derived in \citet{Fields_02} for some SN explosions including furious and normal  SNe and found that the energy distributions of ejecta from the numerical calculations and the empirical formula agree only in the high energy tail when the explosion energy per unit ejecta mass significantly exceeds $1.3\times 10^{51}\mbox{ ergs/}\Msun$. Otherwise, the empirical formula overestimates or underestimates the ejecta mass at around the threshold energy for spallation reactions. Therefore it is necessary to numerically calculate SN explosions to obtain a correct energy distribution of ejecta for estimations of the yield of light elements.

To obtain the yields of light elements from the calculated SN ejecta, we have numerically solved the transfer equation taking into account the energy loss due to ionization of the ISM and spallation reactions with the ISM. The results suggest that light elements synthesized from energetic SNe Ic like SN 1998bw by this mechanism can explain the enrichment of Be and B observed in the Galactic halo stars if the fraction of SN 1998bw like SNe in the early Galaxy was a factor of $>100$ higher than current observational data  suggest. This mechanism must not be the only primary mechanism that worked in the Galactic halo. Other primary mechanisms like superbubbles that supply light elements regardless of SN type are required to reproduce the observed abundance ratios such as Be/Fe. These SNe Ic can produce Li with more than one order of magnitude smaller amounts than indicated by observations. However, lack of information on the Li abundances in the metallicity range of $-2\ltsim$[Fe/H]$\ltsim -1$ prevents us from deducing a firm conclusion. 

SNe are suggested to be associated with aspherical explosions. The deviation from spherical symmetry will be able to increase the mass of ejecta with enough energies for spallation reactions for a given $E_{\rm ex}/M_{\rm ej}$ because $M(>\epsilon)$ is proportional to the $\sim3.4$---3.6 power of this value. To illustrate this effect, a simplified situation will be considered. Suppose the energy injected in the direction with a solid angle of $\pi$ steradian is enhanced by a factor of two and the energy in the other directions is reduced by a factor of $1.5$, then the empirical formula for the mass $M(>\epsilon)$ indicates that this mass will increase by a factor of $\sim 2^{3.4}\times 1/4 + (2/3)^{3.4}\times3/4\sim 2.8$ while the total energy will be unchanged. Applying the empirical formula obtained from the spherically symmetric calculations to this situation might lead to an erroneous result. Thus to further explore SNe Ic as a production site for light elements, we need to perform multi-dimensional relativistic hydrodynamic calculations for SNe Ic that can trace the motion of the outermost ejecta with a sufficient accuracy such as the calculations presented here.


\begin{thebibliography}{}

\bibitem[Blinnikov et al.(2000)]{Blinnikov_00}
 Blinnikov, S., Lundqvist, P., Bartunov, O., Nomoto, K., \& Iwamoto, K.\ 2000, \apj, 532, 1132 

\bibitem[Boesgaard et al.(1999)]{Boesgaard_99} Boesgaard, A.~M., 
Deliyannis, C.~P., King, J.~R., Ryan, S.~G., Vogt, S.~S., \& Beers, T.~C.\ 
1999, \aj, 117, 1549 

\bibitem[Cayrel et al.(2004)]{Cayrel_04} Cayrel, R., et al.\ 
2004, \aap, 416, 1117 

\bibitem[Chaussidon \& Robert(1995)]{Chaussidon_95} Chaussidon, M.~\& 
Robert, F.\ 1995, \nat, 374, 337 

\bibitem[Duncan, Lambert, \& Lemke(1992)]{Duncan_92} 
Duncan,  D.~K., Lambert, D.~L., \& Lemke, M.\ 1992, \apj, 401, 584 

\bibitem[Ensman \& Burrows(1992)]{Ensman_92}
Ensman, L. \& Burrows, A. 1992, \apj, 393, 742

\bibitem[Fields(1996)]{Fields_96}
Fields, B.~D.\ 1996, \apj, 456, 478 

\bibitem[Fields, Olive, Vangioni-Flam, \& Cass{\' e}(2000)]{Fields_00} 
Fields, B.~D., Olive, K.~A., Vangioni-Flam, E., \& Cass{\' e}, M.\ 2000, \apj, 540, 930 

\bibitem[Fields et al.(2002)]{Fields_02} 
Fields, B.~D., Daigne, F., Cass{\' e}, M., \& Vangioni-Flam, E.\ 2002, \apj, 581, 389 

\bibitem[Galama et al(1998)]{Galama_98}
 Galama, T.~J.~et al.\, 1998, \nat, 395, 670

\bibitem[Gandel'man \& Frank-Kamenetskii(1956)]{Gandel'man_56}
Gandel'man, G. M., \& Frank-Kamenetskii, D. A.\, 1956, Soviet-Phys. (Doklady), 1, 223

\bibitem[Higdon, Lingenfelter, \& Ramaty(1998)]{Higdon_98} 
Higdon, J.~C., Lingenfelter, R.~E., \& Ramaty, R.\ 1998, \apjl, 509, L33 

\bibitem[Iwamoto et al.(1994)]{Iwamoto_94}
Iwamoto, K., Nomoto, K., H\"oflich, P., Yamaoka, H., Kumagai, S., \& Shigeyama, T.\ 1994, \apjl, 
437, L115 

\bibitem[Kazhdan \& Murzina(1992)]{Kazhdan_92}
Kazhdan, Ya., M. \& Murzina, M. 1992, \apj, 400, 192

\bibitem[Kulkarni et al.(1998)]{Kulkarni_98}
Kulkarni, S.~R.~et al.\ 1998, \nat, 395, 663 

\bibitem[Lambert et al.(1998)]{Lambert_98} Lambert, D.~L., 
Sheffer, Y., Federman, S.~R., Cardelli, J.~A., Sofia, U.~J., \& Knauth, 
D.~C.\ 1998, \apj, 494, 614 

\bibitem[Maeda et al.(2002)]{Maeda_02} Maeda, K., Nakamura, T., 
Nomoto, K., Mazzali, P.~A., Patat, F., \& Hachisu, I.\ 2002, \apj, 565, 405 

\bibitem[Mart{\'{\i}} \& M{\" u}ller(1996)]{Marti_96}
Mart{\'{\i}}, J.~M., \& M{\" u}ller, E. 1996, J.~Comput.~Phys., 123, 1

\bibitem[Matzner \& McKee(1999)]{Matzner_99}
Matzner, C.~D.~\& McKee, C.~F.\ 1999, \apj, 510, 379 

\bibitem[Mazzali et al.(2002)]{Mazzali_02} Mazzali, P.~A.~et al.\ 
2002, \apjl, 572, L61 

\bibitem[McWilliam et  al.(1995)]{McWilliam_95} 
McWilliam, A., Preston, G.~W., Sneden, C., \& Searle, L.\ 1995, \aj, 109, 2757 

\bibitem[Meneguzzi, Audouze, \& Reeves(1971)]{Meneguzzi_71} 
Meneguzzi, M., Audouze, J., \& Reeves, H.\ 1971, \aap, 15, 337 

\bibitem[Nakamura et al.(2001)]{Nakamura_01} 
Nakamura, T., Mazzali, P.~A., Nomoto, K., \& Iwamoto, K.\ 2001, \apj, 550, 
991 

\bibitem[Nomoto, Filippenko, \& Shigeyama(1990)]{Nomoto_90}
Nomoto, K., Filippenko, A. V., \& Shigeyama, T. 1990, \aap, 240, L1

\bibitem[Podsiadlowski et al.(2004)]{Podsiadlowski_04}
Podsiadlowski, Ph., Mazzali, P. A., Nomoto, K., Lazzati, D., \& Cappellaro, E.\ 2004, \apjl, in press

\bibitem[Read \& Viola(1984)]{Read_84} 
Read, S.~M.~\& Viola, V.~E.\ 1984, Atomic Data and Nuclear Data Tables, 31, 359 

\bibitem[Ryan, Norris, \& Beers(1999)]{Ryan_99} Ryan, S.~G., 
Norris, J.~E., \& Beers, T.~C.\ 1999, \apj, 523, 654 

\bibitem[Ryan et al.(2000)]{Ryan_00} Ryan, S.~G., Beers, T.~C., 
Olive, K.~A., Fields, B.~D., \& Norris, J.~E.\ 2000, \apjl, 530, L57 

\bibitem[Ramaty \& Lingenfelter(1975)]{Ramaty_75} Ramaty, R.~\& 
Lingenfelter, R.~E.\ 1975, IAU Symp.~ 68: Solar Gamma-, X-, and EUV 
Radiation, 68, 363

\bibitem[Ramaty et al.(2000)]{Ramaty_00}
 Ramaty, R., Scully, S.~T., Lingenfelter, R.~E., \& Kozlovsky, B.\ 2000, \apj, 534, 747

\bibitem[Sakurai(1960)]{Sakurai_60}
Sakurai, A. 1960, Comm. Pure Appl. Math. 13, 353

\bibitem[Schlickeiser(2002)]{Schlickeiser_02} Schlickeiser, R.\ 2002, 
in Cosmic ray astrophysics / Reinhard Schlickeiser, Astronomy and Astrophysics 
Library; Physics and Astronomy Online Library.~Berlin: Springer.~ISBN 
3-540-66465-3, 2002, XV

\bibitem[Shigeyama \& Nomoto(1990)]{Shigeyama_90}
Shigeyama, T.~\& Nomoto, K.\ 1990, \apj, 360, 242 

\bibitem[Shigeyama \& Tsujimoto(1998)]{Shigeyama_98}
Shigeyama, T. \& Tsujimoto, T. 1998, \apj, 507, L135

\bibitem[Shima(1962)]{Shima_62} Shima, M.\ 1962, \jgr, 67, 4521 

\bibitem[Spitzer(1978)]{Spitzer} Spitzer, L.\ 1978, New York 
Wiley-Interscience, 1978.~333 p.,  

\bibitem[Suzuki \& Yoshii(2001)]{Suzuki_01}
 Suzuki, T.~K.~\& Yoshii, Y.\ 2001, \apj, 549, 303

\bibitem[Tan, Matzner, \& McKee(2001)]{Tan_01}
Tan, J. C., Matzner, C. D., \& McKee, C. F. 2001, \apj, 551, 946

\bibitem[Tsujimoto et al.(1997)]{Tsujimoto_97}
Tsujimoto, T., Yoshii, Y., Nomoto, K., Matteucci, F., Thielemann,
F.-K., \& Hashimoto, M. 1997, ApJ, 483, 228

\bibitem[Tsujimoto, Shigeyama, \& Yoshii(1999)]{Tsujimoto_99}
Tsujimoto, T., Shigeyama, T., \& Yoshii, Y.\ 1999, \apjl, 519, L63

\end{thebibliography}
\end{document}